\documentclass[12pt]{article} 
\usepackage{graphicx}
\usepackage{amsmath}
\usepackage{amssymb}
\usepackage{units}
\usepackage{xcolor}
\usepackage[numbers, sort&compress, elide]{natbib}

\usepackage{jheppub}

\def\equationautorefname~#1\null{Eq.\,(#1)\null}

\makeatletter\g@addto@macro\bfseries{\boldmath}\makeatother

\newcommand{\bea}{\begin{eqnarray}}
\newcommand{\eea}{\end{eqnarray}}
\newcommand{\beq}{\begin{equation}}
\newcommand{\eeq}{\end{equation}}

\def\Cambridge{DAMTP, University of Cambridge, Wilberforce Road, Cambridge, CB3 0WA, UK}

\usepackage[normalem]{ulem}

\begin{document}

\title{Gravitational Wave Backgrounds from Colliding ECOs}

\author[a,b]{Hannah Banks}
\author[a,c]{, Dorota M. Grabowska}
\author[a]{and Matthew McCullough}

\affiliation[a]{Theoretical Physics Department, CERN, 1211 Geneva 23, Switzerland}
\affiliation[b]{\Cambridge}
\affiliation[c]{InQubator for Quantum Simulation (IQuS), Department of Physics, University of Washington, Seattle, WA 98195}

\emailAdd{hmb61@cam.ac.uk}
\emailAdd{grabow@uw.edu}
\emailAdd{matthew.mccullough@cern.ch}

\preprint{CERN-TH-2023-016 
\begin{flushright} IQuS@UW-21-044
\end{flushright}}

\abstract{Long baseline atom interferometers offer an exciting opportunity to explore mid-frequency gravitational waves.  In this work we survey the landscape of possible contributions to the total `gravitational wave background'  in this frequency band and advocate for targeting this observable.  Such an approach is complimentary to searches for resolved mergers from individual sources and may have much to reveal about the Universe. We find that the inspiral phases of stellar-mass compact binaries cumulatively produce a signal well within reach of the proposed AION-km and AEDGE experiments.  Hypothetical populations of dark sector exotic compact objects, harbouring just a tiny fraction of the dark energy density, could also generate signatures unique to mid- and low-frequency gravitational wave detectors, providing a novel means to probe complexity in the dark sector.}

\maketitle

\section{Introduction}

The pursuit for fundamental physics beyond the Standard Model (SM) is entering a new era.  It is becoming increasingly attractive to explore the possibility that the solutions we seek have remained hidden not behind a currently unattainable energy barrier, but through incredibly weak couplings to the SM.  Recent developments in quantum sensing technologies have unlocked a host of new avenues through which to probe this `feebly interacting frontier' (see e.g.~\cite{Lanfranchi:2020crw}). By measuring the phase difference between atomic matter waves travelling along  different paths, atom interferometry is one such class of experiment that offers exciting opportunities to probe physics at the quantum level.  By coupling state-of-the-art developments in the technologies deployed in atomic clocks with established methods of constructing inertial sensors, it has become feasible to consider upscaling conventional table-top experiments in order to reach  lower operational frequencies. Five long-baseline terrestrial prototype atom interferometry experiments, AION-10  \cite{Badurina:2019hst} in the UK  , MAGIS-100 \cite{Abe_2021} in the US , MIGA  \cite{Canuel_2018} in France, VLBAI \cite{Hartwig_2015}  in Germany and ZAIGA  \cite{Zhan_2019} in China, have currently been commissioned and are under construction, serving as essential technology readiness indicators to the deployment of km-scale terrestrial detectors by the mid 2030s. In addition to providing opportunities to search for exotic forces and hidden sectors of particles, experiments like AION-km, MAGIA-advanced, ELGAR \cite{Canuel_2020}, MAGIS-km and the advanced-ZAIGA, will open a unique window to the mid-frequency gravitational wave (GW) landscape. In particular, due to the specific frequency bands that they can probe, they  bridge the gap in frequencies that will exist between the  LIGO and LISA detectors. Proposals for space-based cold-atom interferometers, most notably AEDGE \cite{AEDGE:2019nxb}, are also under serious consideration, and would allow for an even more complete exploration of this unchartered phenomenological territory.  For the purpose of example, this work will be primarily concerned with the AION-km experiment, and its possible successor AEDGE. Although to be realised using different cold-atom technologies, with predicted sensitivities of a similar order of magnitude, we expect the conclusions reached in this work to also be of relevance to other km-scale atom interferometry experiments. 

To date, the science cases for the GW programmes of km-scale atom interferometers have largely, but not entirely, focused on the possibility of detecting resolved mergers of individual  intermediate-mass black hole binary systems. There have also been proposals to search for stochastic signals arising from early Universe first-order phase transitions or cosmic strings. The prospective sensitivities of AION-km and AEDGE to such signals are considered in Refs.~\cite{Badurina:2019hst} and \cite{AEDGE:2019nxb} and explored more comprehensively in Ref.~\cite{Badurina:2021rgt}. In this work we examine the AION capabilities through an alternative lens, turning instead to the cumulative signals or `gravitational wave backgrounds' (GWB) that arise from the incoherent superposition of the radiation from large numbers of, for the most part, individually unresolvable merging binary systems.  Whilst their peak emission occurs at the point of merger in the LIGO frequency band, stellar-mass astrophysical compact binary systems produce a continuous spectrum of lower frequency gravitational radiation during their inspirals. Using the distributions and merger rates of these populations as inferred from direct observations at the  LIGO-Virgo network, we demonstrate that the cumulative background from these LIGO-Virgo populations of  compact binaries is well within the experimental reach of both terrestrial, and space-based long-baseline atom interferometers. As well as being a relevant and previously unexplored background that needs to be mapped out in order to assess the sensitivity of atom interferometers to other, perhaps more exotic, sources of GW, measurements of this signal itself could have much to tell us.

Several calculations of the backgrounds from mergers of the LIGO astrophysical populations  have been presented previously in literature (see, for example, Refs.~\cite{2020, Lewicki:2021kmu, KAGRA:2021kbb}); however their specific implication for the mid-frequency band terrestrial long-baseline atom interferometry has received little attention. In Section \ref{general} we establish a general procedure for calculating the total gravitational background signal from a given population of merging compact objects. This treatment is largely based on Ref.~\cite{Lewicki:2021kmu}, which computed the stochastic background from the stellar-mass binary black hole (BBH) population inferred from LIGO-Virgo observations in order to establish its implications for future measurements of sub-dominant cosmological backgrounds. In Sections \ref{BBH}, \ref{BNS} and \ref{BHNS} we apply this framework specifically to the populations of stellar-mass black holes (BBH), neutron stars (BNS), and black hole-neutron star (BHNS) binary systems that have been inferred from the LIGO-Virgo observations in order to estimate their total astrophysical signal in the AION frequency band.  In addition to forming a relevant background to more exotic sources of GW, we argue that this signal offers an interesting science case from both an astrophysical and fundamental physics perspective, meriting close attention in its own right.   Although we expect these sources to dominate the total astrophysical background for terrestrial based long-baseline atom interferometers, in Section \ref{other} we comment on alternative sources that may be of relevance, particularly at the lower frequencies that are to be reached in future space-based experiments. However, these alternative sources have greater uncertainty in the estimation of their contribution to the total gravitational background signal;  this is due to the lack of empirical data that informs the estimation of distribution and merger rates.  As such we do not include them in our estimation for the total astrophysical background.

Given the rich landscape of astrophysical structures present in the vastly subdominant visible sector of the Universe, it is reasonable to assume that the dark sector may harbour similar diversity.  Indeed, the coalescence of dark particles or exotic states of Standard Model particles into a stable astrophysical-size   `Exotic Compact Objects'  (ECOs) has been a subject of much recent attention in literature \cite{Macedo:2013jja,Giudice:2016zpa,Cardoso:2016oxy}. Examples of ECOs are fermion stars~\cite{Itoh:1970uw, Bodmer:1971we, Witten:1984rs, Alcock:1986hz, Madsen:1998uh, Weber:2004kj, Narain:2006kx},  boson stars \cite{Jetzer:1991jr, Liddle:1992fmk, Schunck:2003kk, Liebling:2012fv, Visinelli:2021uve, RevModPhys.91.041002}, and dark matter stars \cite{Kouvaris:2015rea}. These ECOs fit into a paradigm for dark matter where elementary particles form exotic composite objects with exponentially large occupation numbers.  Since the mass of these objects span many decades, they must be searched for in a multitude of ways, including both terrestrial and non-terrestrial methods \cite{Gresham:2017zqi, Gresham:2017cvl, Grabowska:2018lnd,  Coskuner:2018are, Bai:2018dxf, Bhoonah:2020dzs, Das:2021drz, Acevedo:2021kly,  Baum:2022duc, Bai:2022nsv}. Returning to ECOs specifically, if these objects form binary systems, the gravitational radiation resulting from their mergers would contribute to the overall background, much in the same way as their SM counterparts, and could potentially be observable. Estimates of the  background signals from a handful of specific types of binary ECOs, namely boson stars and dark blobs, are considered in Ref.~\cite{Croon_2018} and Ref.~\cite{Diamond:2021dth} respectively.  Here, we attempt to take a more general approach, keeping our analysis as model-independent as we can, and classifying signals only in terms of the mass scale of the binaries involved. In Section \ref{ECOS} we show that even if such populations only comprise a very small fraction of the dark matter budget, they could generate detectable signals in the mid-range frequency band to be probed by long baseline atom interferometers. Combining measurements of the cumulative background at experiments operating in complimentary frequency ranges, this signal could be distinguishable from the dominant SM  background, giving long-baseline atom interferometers a unique capability to probe the possible existence, formation history and distribution of these structures, and hence complexity in the dark sector.

\section{ GW backgrounds from Compact Binary Mergers }
\label{general}
We shall begin by clarifying exactly what we mean by a gravitational wave background (GWB).  Although just a small fraction of the rich and diverse GW landscape, the only direct detections of gravitational radiation to-date have been from isolated point-like sources that generate coherent, resolvable signals. In practice however, many signals that reach any given GW detector cannot be resolved. This occurs when individual signals are too weak or when they are too closely spaced in time to be separated  \cite{Renzini:2022alw}. The total gravitational emission from a given population of sources, including both resolvable and unresolvable signals, is an intrinsic property of the Universe \cite{Abbott_2016} and forms what we shall refer to as the GWB for that source type. It is this, summed over all possible source types, that may be measured by the detector. If the resolvable signals are subsequently removed,  the remaining `residual' background, composed from the incoherent accumulation of individually unresolved signals, can be analysed stochastically. This stochastic gravitational wave background (SGWB) is thus detector-dependent and can receive both astrophysical and cosmological (primordial) contributions \cite{Rosado:2011kv}. The cosmological background is vastly subdominant, and much thought has been given to determine ways of distinguishing it from the astrophysical stochastic background; see for example Ref. \cite{Martinovic_2021}.  A comprehensive, recent review of the SGWB and its various SM sources can be found in Ref.~\cite{Renzini:2022alw}.  In reality, the practice of removing resolved signals is far from straightforward, as it is detector-dependent and involves a number of subtleties \cite{Renzini:2022alw}. Given this, and accounting for the fact that in this work we shall consider several different detectors of which the experimental details are not yet certain, we shall only concern ourselves with the total GWB and not attempt to exclude potentially resolvable sources from our computations.\footnote{We note that some works refer to any cumulative calculation of gravitational wave emission from a set of sources as a contribution to the SGWB even if resolved signals are not excluded. Whilst this may be practical given the large uncertainties and degree of choice involved in any subtraction procedure, for the background to be stochastic in a statistical sense, any potentially resolvable astrophysical signal should be removed. This distinction of terminology is only relevant for the calculation of cumulative signals from populations of objects whose individual signals are not intrinsically stochastic, not where the sources are inherently stochastic such as those of cosmological origin \cite{Ginat_2020,Renzini:2022alw}.}

We shall henceforth review the formalism for computing the GWB from a generic, known, population of compact binary systems. Throughout we shall adopt the convention that the speed of light, $c$ = 1, assume a flat standard $\Lambda$CDM cosmology with $\Omega_m$ = 0.3065 and $\Omega_\Lambda$ = 1 - $\Omega_m$ , and  take the Hubble constant, $H_0$,  to be 67.9 km/s/Mpc, as measured by the Planck collaboration \cite{2016A&A...594A..13P}.  

GWBs can be conveniently characterised in terms of their energy density spectrum via the dimensionless quantity 
\begin{equation}
\Omega_{GW}(f) = \frac{f}{\rho_c}\frac{\textnormal{d}\rho_{GW}(f)}{\textnormal{d} f}~,
\end{equation}
where $\textnormal{d}\rho_{GW}$   is the total energy density in the observed frequency range $f$ to $f+\textnormal{d}f$, and $\rho_c = 3H_0^2/(8 \pi G) $ is the critical density needed to close the Universe. 

If one specifies to the  GWB from mergers of a population of binary compact objects with constituent masses $m_1$ and $m_2$ located at a redshift $z$, this can be further expressed as
\begin{equation}
\Omega_{GW}(f) = \frac{f}{\rho_c H_0} \int_0^{z_{\textnormal{max}}} \textnormal{d}z \int \textnormal{d}m_1 dm_2 \frac{\textnormal{d}\mathcal{R}}{\textnormal{d}m_1 \textnormal{d}m_2}\frac{\textnormal{d}\tilde{E}_{\textnormal{GW}}}{\textnormal{d}f_s}\frac{1}{(1+z)E(z)}~.
\end{equation}
Here, $f_s$ refers to the emission frequency in the source frame, which is related to $f$, the frequency measured at the detector, by $f_s = f(1+z)$. Additionally, $\frac{\mathcal{R}(z)}{dm_1 dm_2}$ is the differential rate of mergers per comoving volume element and $\frac{\textnormal{d}\tilde{E}_{\textnormal{GW}}}{\textnormal{d}f_s}$ is the source frame energy spectrum from a single binary system. The function $E(z)$ is defined as $\sqrt{\Omega_m (1+z)^3 + \Omega_\Lambda }$.

This expression can be recast as
 \begin{equation}
\Omega_{\textnormal{GW}} = \int \textnormal{d}m_1 \textnormal{d}m_2 \int \frac{\textnormal{d}V_c}{1+z} \frac{\textnormal{d}\mathcal{R}(z)}{\textnormal{d}m_1 \textnormal{d}m_2 } \frac{1}{\rho_c} \frac{\textnormal{d}\tilde{\rho}_{\textnormal{GW}}(m_1,m_2)}{df}~,
\end{equation}
where  $\frac{\textnormal{d}\tilde{\rho}_{\textnormal{GW}}}{\textnormal{d}f}$ is the comoving energy density spectrum of a single binary system in the detector frame, and $\textnormal{d}V_c $ is the comoving volume element. The former is directly related to the amplitude of the Fourier transform of the detected signal, $|\tilde{h}(f)|$ according to \cite{Moore_2014}
\begin{equation}
d\tilde{\rho}_{\textnormal{GW}} = \frac{4}{5}\frac{\pi}{G}f^{3}|\tilde{h}(f)|^2 df~,
\end{equation} where the factor of 4/5  accounts for averaging over possible source orientations \cite{Zhu:2012xw,Sathyaprakash:2009xs}. 

Although $|\tilde{h}(f)|$ is specific to the class of objects merging, and is in general not analytically tractable, it is worth pointing out that the dominant contribution to the background comes from the inspiral phase of the merger. Here, the binary constituents are widely separated and the system can be well-approximated by a pair of self-gravitating point masses emitting gravitational quadrupole radiation. This calculation is both analytic and irrespective of the specific nature of the objects involved and results in a contribution to $\Omega_{GW}$ that scales as $f^{2/3}$ \cite{Peters:1963ux}.

The final component of the calculation is the differential merger rate, $\frac{\mathcal{R}(z)}{dm_1 dm_2}$, which encodes both the redshift, and mass distributions of the population in question.  Taking these distributions to be decoupled,\footnote{This can be assumed without loss of generality: the dominant contribution to the GWB  comes from nearby sources meaning that the calculation is most impacted by the value of the matter distribution locally. The exact functional form of the redshift distribution is of lesser importance.} this can expressed as
\begin{equation}
\frac{\textnormal{d}R(t)}{\textnormal{d}m_{1}\textnormal{d}m_{2}} =  \frac{R_0}{Z} P_m(m_1,m_2) \int \textnormal{d}t_{d} dz_{b} P_b (z_b)P_d(t_d)\delta(t - t_d - t(z_b))~,
\end{equation} where $P_m(m_1, m_2)$ is a function of the constituent masses that encodes the mass distribution, $R_0$ is the current merger rate, and $Z$ is a normalisation constant chosen such that 
\begin{equation}
R(z=0) = \int dm_1 dm_2 \left. \frac{dR}{dm_1 dm_2} \right\rvert_{z=0} \equiv R_0~.
\end{equation}
The redshift distribution is typically taken as the convolution of the binary formation rate, $P_b(z_b)$ with the distribution of time delays $P_d(t_d)$ between formation and merger.  Each of these distributions are specific to the population in question. Armed with this general prescription, we shall now consider a number of distinct astrophysical populations:  stellar-mass binary black holes (BBH), binary neutron stars (BNS) and black hole - neutron star binary systems (BHNS), before eventually casting our attention in Section \ref{ECOS} to hypothetical populations of ECOs. 

\subsection{Binary Black Holes}
\label{BBH}
 Thanks to the (now sizeable) number of direct merger observations made by the LIGO-Virgo network \cite{LIGOScientific:2018mvr,  LIGOScientific:2020ibl,  LIGOScientific:2021djp}, the distribution and merger rate of the population of $\mathcal{O}(10M_{\odot})$ binary black holes are well understood. In this way, and in contrast to the increasingly elusive and eventually hypothetical systems to which we shall later turn, calculations of the cumulative GW signal from these BBH mergers are empirically informed.  Recent observations suggest that the population of stellar-mass BBHs probed by the LIGO-Virgo network have small effective spins \cite{LIGOScientific:2020kqk}. As such, we follow Ref.~ \cite{KAGRA:2021kbb} in assuming the BBH spins to be negligible. 
 
Taking this population to be purely astrophysical, with no primordial contribution, the  BH formation rate can be assumed to track the stellar formation rate (SFR), which we take to be \cite{Madau_2017} 
\begin{equation}
\label{SFR}
SFR(z) \propto \frac{(1+z)^{2.6}}{1+((1+z)/3.2)^{6.2}} \equiv P_b (z) ~.
\end{equation} We assume the formation-merger delay time has a simple power law distribution $P_d(t_d) \propto t_d^{-1}$ over $t_{min} < t_d < t_{max}$ where $t_{min}$ = 50 Myr and $t_{max}$  is the age of the Universe at merger, $t(z)$ \cite{Belczynski:2016obo}. 

We adopt the model for the BBH mass distribution introduced in Ref.~\cite{H_tsi_2021} and employed in Ref.~\cite{Lewicki:2021kmu},
 \begin{equation}
P_m(m_1,m_2) =   M^{\alpha} \eta^{\beta} \psi(m_1)\psi(m_2)~~.
\end{equation}

Within this expression, $M = m_1 + m_2$ is the total mass of the system and $\eta = m_1 m_2 / (m_1+m_2)^2$, is the symmetric mass ratio. The mass function of the constituent objects is given by $\psi$ and is taken to be a truncated power law of the form 
\begin{equation}
\psi(m) \propto m^{\zeta} \Theta(m - m_{\textnormal{min}}) \Theta(m_{\textnormal{max}} - m)~,
\end{equation} 
where $\Theta$ denotes the heaviside step function and the normalisation is such that
\begin{equation}
\int  \psi(m) d\ln m=1~~.
\end{equation}
The minimum and maximum masses within this function are taken to be $m_{\textnormal{min}} = 3 M_{\odot}$ and  $m_{\textnormal{max}} = 55 M_{\odot} $, corresponding to estimates for the maximum mass of neutron stars and the start of the pair instability mass gap respectively \cite{Lewicki:2021kmu}. The mass dependence of astrophysical binary system formation is empirically encoded via the parameters $\alpha$ and $\beta$. Following the maximum likelihood fit to the LIGO-Virgo event catalogue in Ref.~\cite{H_tsi_2021} we adopt the values  $\alpha $ = 0, $\beta $ = 6, $\zeta = -1.5$ and $R_0 = 10^{+6}_{-5}$ Gpc$^{-3}$yr$^{-1}$.

For BBHs, the Fourier transformed signal amplitude for inspiral, merger and ringdown phases can be respectively  approximated as
\begin{equation}
\label{eq:wf}
|\tilde{h}(f)| = \sqrt{\frac{5\eta}{24}} \frac{[GM(1+z)]^{5/6}}{\pi^{2/3}d_L} \times \begin{cases} f^{-7/6} & f <  f_{\textnormal{merg}} \\
f_{\textnormal{merg}}^{-1/2}f^{-2/3} & f_{\textnormal{merg}} \leq f < f_{\textnormal{ring}} \\
f_{\textnormal{merg}}^{-1/2}f_{\textnormal{ring}}^{-2/3}\frac{\sigma^2}{4(f - f_{\textnormal{ring}})^2 + \sigma^2} & f_{\textnormal{ring}} \leq f < f_{\textnormal{cut}} \\
0 & f \geq f_{\textnormal{cut}}
\end{cases}.
\end{equation}
The frequencies $f_{\textnormal{merge}}, f_{\textnormal{ring}}, f_{\textnormal{cut}}$ and the ringdown width $\sigma$ are of the form
\begin{equation}
f_j = \frac{a_j \eta^2 + b_j \eta + c_j}{\pi G M (1+z)}~,
\end{equation}
with the coefficients $a_j$, $b_j$ and $c_j$ as listed in Ref.~\cite{Ajith:2007kx}. $d_L(z) = (1+z)/H_0 \int_0^z E(z')^{-1} dz' $ is the luminosity distance to the source and is related to the comoving distance $d_c(z)$ by $d_L(z) = (1+z) d_c(z)$.

\subsection{Binary Neutron Stars}
\label{BNS}
With only two recorded direct observations of mergers to date  \cite{LIGOScientific:2017vwq,LIGOScientific:2020aai}, there remains considerable uncertainty in the mass and redshift distributions of binary neutron stars and, in turn, the determination of their present event rate.  Following the analysis in Refs.~\cite{LIGOScientific:2019vic,KAGRA:2021kbb}, we take each of the component neutron stars to follow a uniform mass distribution over the range 1 $M_{\odot}$-2.5 $M_{\odot}$. We take the BNS progenitor formation rate to be proportional to that of stellar formation as given in Eq.~\ref{SFR} and once again assume a time delay distribution of the form $P_d(t_d) \propto t_d^{-1}$ for $t_{min} < t_d < t_{max}$ where in this case $t_{min}$ = 20 Myr and $t_{max}$  is the age of the Universe at merger, $t(z)$ \cite{LIGOScientific:2019vic,KAGRA:2021kbb}. We adopt the value of the present event rate given in Ref.~\cite{KAGRA:2021kbb} of $R_0$ = $320\pm^{490}_{240}$ Gpc$^{-3}$ yr$^{-1}$. This estimation was based on an alternative parametrisation of the redshift distribution compared to that deployed in our calculations, specifically in the form of the stellar formation rate. Whilst this choice could, in principal, have impacted the value of $R_0$ obtained in the fit,  given the magnitude of the uncertainties involved, we consider this determination  sufficient for our purposes. Due to the lack of observational data, the waveforms of the post-inspiral phases of binary NS merger events are not well understood. As such, we follow Ref.~\cite{KAGRA:2021kbb} in only including contributions to $\Omega_{BNS}$ from the inspiral phase, truncating the waveform at the frequency corresponding to the innermost stable circular orbit (ISCO) at which stage the separation of the merging objects renders a point-particle approximation invalid.  For a binary system, the ISCO frequency,\footnote{Note that this formula is the ISCO formula specifically for black holes. More rigorously one should account for object compactness, $C$, as in  Eq.~\ref{isco}. Here, we use the black-hole compactness in order to match the approach taken by the LIGO collaboration in Ref.~\cite{KAGRA:2021kbb}. We note that this provides a reasonable cut-off scale for this purpose and acts in the direction of balancing out the neglected contributions from post-inspiral phases.} $f_\textnormal{I}$, (at the source) is \cite{Giudice:2016zpa}
\begin{equation}
\label{fisco}
f_{\textnormal{I}}   = \frac{1}{6^{3/2}\pi GM}~,
\end{equation}
where as before $M$ refers to the total mass of the binary system.

\subsection{Black Hole-Neutron Star}
\label{BHNS}
Whilst no direct observations of binary neutron star - black hole mergers have been confirmed by the LIGO-Virgo network at present \cite{KAGRA:2021kbb}, there exist two possible event candidates \cite{LIGOScientific:2020ibl,LIGOScientific:2020zkf}. While the actual physical nature of these events remain undetermined at this time, we use them to construct a conservative upper bound on the population of BHNS. In the absence of empirical data to ground models of these events, we follow Ref.~\cite{KAGRA:2021kbb} in assuming the BH and NS to follow independent delta-function mass distributions at 10 $M_{\odot}$ and 1.4 $M_{\odot}$ respectively, and adopt an identical redshift distribution to that used previously for BBH mergers. Being an upper estimate, we include contributions to the energy spectra from each of the inspiral, merger and ringdown phases, using the BBH amplitudes given in Eq. \ref{eq:wf}. We emphasise that since some proportion of BHNS inspirals are likely to terminate with tidal-disruption of the neutron star \cite{PhysRevD.83.024005,PhysRevD.92.024014, PhysRevD.82.044049}, the inclusion of these contributions likely overestimates $\Omega_{BHNS}$ at high frequencies. Given the operational range of the experiments considered in this work, this does not pose a problem to our analysis. We take the present day merger rate to be $3.1 \times 10^3$ Gpc$^{-3}$yr$^{-1}$ which is the upper limit obtained in the analysis of Ref. \cite{LIGOScientific:2018mvr} and used by both Ref.~\cite{LIGOScientific:2019vic} and Ref.~\cite{KAGRA:2021kbb} . 

\subsection{Backgrounds from other Astrophysical Sources }
\label{other}
Although we expect the GWB from the inspiral of stellar-mass compact binaries to largely dominate the total astrophysical background within the AION mid-frequency band,  for completeness we shall briefly comment on other potential contributions. With the absence of empirical data such as that enjoyed by the LIGO compact binaries, it is important to note that the population characteristics and potential merger rates of the astrophysical populations to be discussed here are highly uncertain, and in some cases completely unknown, making their potential impact difficult to assess. As such, we do not include these sources in our presentations of the total astrophysical GWB in Section \ref{analysis} but shall instead limit ourselves to a predominantly qualitative discussion here. Our treatment largely follows that in Ref.~\cite{Barish_2021}, to which we refer the interested reader for further details. 

We first turn to possible signals from Intermediate-Mass Black Holes (IMBHs), a hypothetical population of BHs with masses between  $\sim 10^2 - 10^4 M_{\odot}$ \cite{Amaro_Seoane_2007}, of which just one indirect observation has been made \cite{LIGOScientific:2020iuh}.  Whilst the formation mechanism of these objects is yet to be established, `runaway' mergers of massive stars in young dense stellar clusters \cite{Ebisuzaki_2001} is one possibility. As such, one may postulate the existence of so-called Intermediate-Mass-Ratio-Inspirals (IMRIs)  which arise from the mergers of stellar-mass and IMBHs. The rate of such mergers would depend on a number of unknowns, not least the IMBH mass and redshift distributions. A broad-brush estimation of the background from such events is presented in Ref.~\cite{Barish_2021}. Here, the IMBHs are chosen to follow a uniform mass distribution, and a merger rate selected according to predictions of the number of IMRIs made in Ref.~\cite{Amaro_Seoane_2007}. The resulting spectrum, which therefore also scales as $f^{2/3}$ during inspiral, is subdominant to that of stellar-mass black holes at all frequencies. Whilst IMBH-IMBH mergers are also possible, these events are likely to be even rarer and therefore also likely to have backgrounds inferior to their stellar-mass counterparts.  We therefore do not explicitly estimate either background.  Mergers between stellar-mass and supermassive black holes, so called Extreme-Mass-Ratio-Inspirals \cite{Amaro_Seoane_2007}, provide another potentially sizeable  background. Modelling these signals is a complex process, not least because there is a large uncertainty in the merger rate and because many of the relevant population parameters are unknown. Whilst the rough estimate of this background made in Ref.~\cite{Barish_2021} places it as subdominant to the stellar-mass black hole background at the frequencies most relevant to AION, the two become comparable in a small frequency range localised around $10^{-3}$ Hz which will be accessible to LISA and AEDGE+. Given that this estimation is subdominant to that expected from stellar-mass compact binaries at the frequencies probed by the terrestrial long-baseline experiments and first-generation space based atom interferometers that form the basis of this work, we do not include it in our total astrophysical background as presented in Sec. \ref{analysis}. Given the high uncertainty in this forecast, we note again that once more detailed data is accrued on EMRIs, a more detailed consideration of this signal is warranted in order to fully assess its relevance. The same can be said for IMRIs. In the meantime however, one must remain mindful of these sources, particularly when attempting to interpret the origin of possible observations of  `exotic' backgrounds.  Finally,  as argued in Ref.~\cite{Barish_2021}, we note that the stochastic backgrounds generated from mergers of binary supermassive black holes are typically not deemed to be relevant for mid-frequency GW experiments, neither are those from slowly rotating neutron stars or type 1a supernovae. 

Although not relevant in the AION frequency band, white dwarf binaries (WDB) become important at the lower frequencies reached by space-based missions including AEDGE and LISA. The emission from these objects is weak and the background is overwhelmingly dominated by binaries from within the galaxy. The stochastic (ie. unresolved) galactic WDB background spectrum, as estimated for LISA, is predicted to exceed that of stellar-mass compact binaries sub $\sim 0.003$ Hz \cite{Robson:2018ifk}. Being highly anisotropic, one should in principle be able to exploit the yearly modulation of this background to remove all but its isotropic component, allowing it to be neglected \cite{Pieroni_2020}.

We thus conclude, in the absence of data on which to form more detailed models, that the other potential astrophysical backgrounds are either largely subdominant to that of stellar-mass binaries or excludable such that to a good approximation they can be neglected in estimations of the total astrophysical background. 

\begin{figure}
\centering
\includegraphics[width=\textwidth]{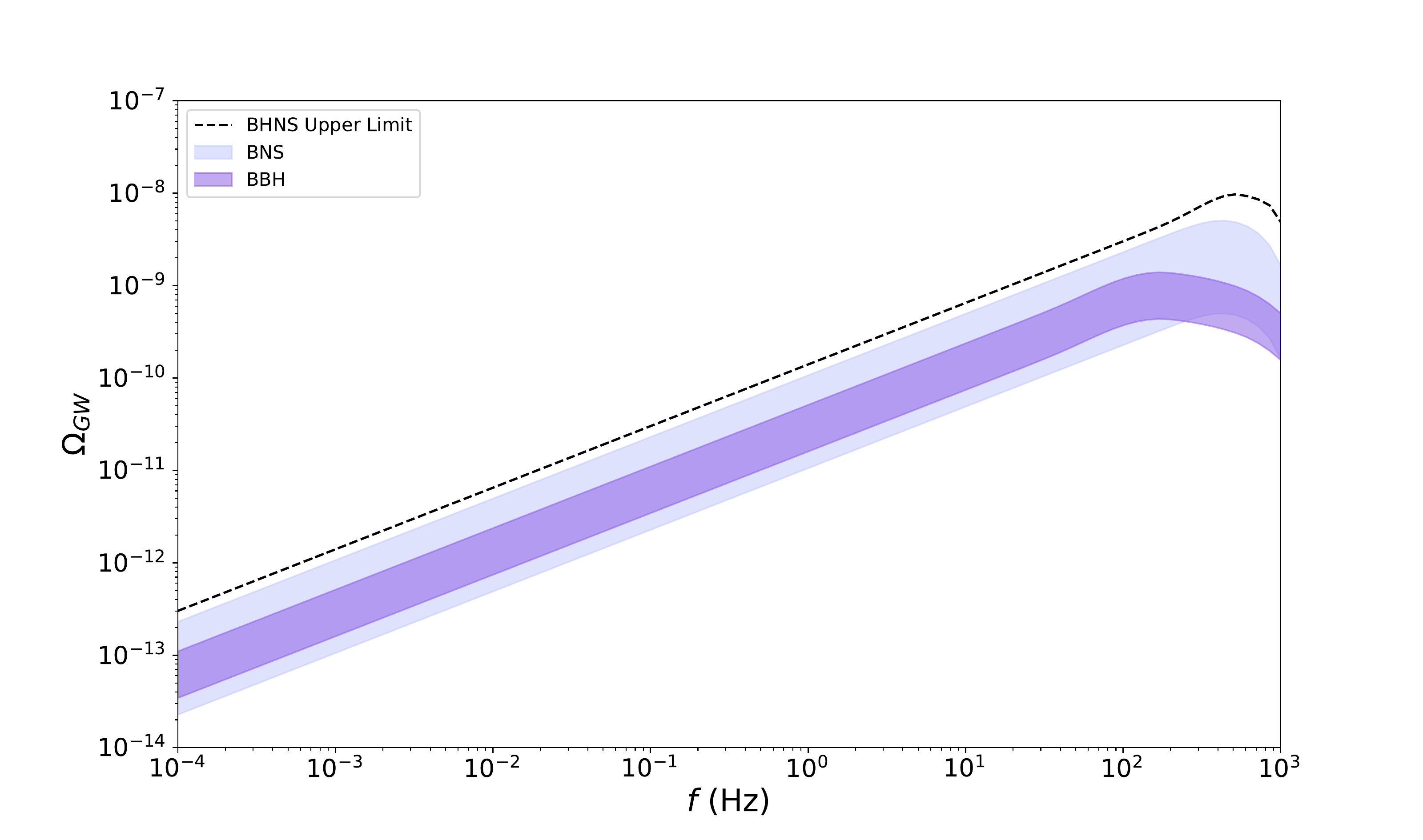}
\caption{Estimates of the  gravitational wave backgrounds from mergers of stellar-mass  BH, NS and BHNS binaries. The signal from BH and NS binary populations are shown as bands whose upper (lower) values arise from adopting the 1 $\sigma$ upper (lower) limits of the present merger rate $R_0$ as extracted from LIGO-Virgo data in Ref.~\cite{H_tsi_2021} and Ref.~\cite{KAGRA:2021kbb}  respectively. The dashed line is a (conservative) upper estimate of the signal from BHNS mergers.}
\label{fig:components}
\end{figure}

\begin{figure}
\centering
\includegraphics[width=\textwidth]{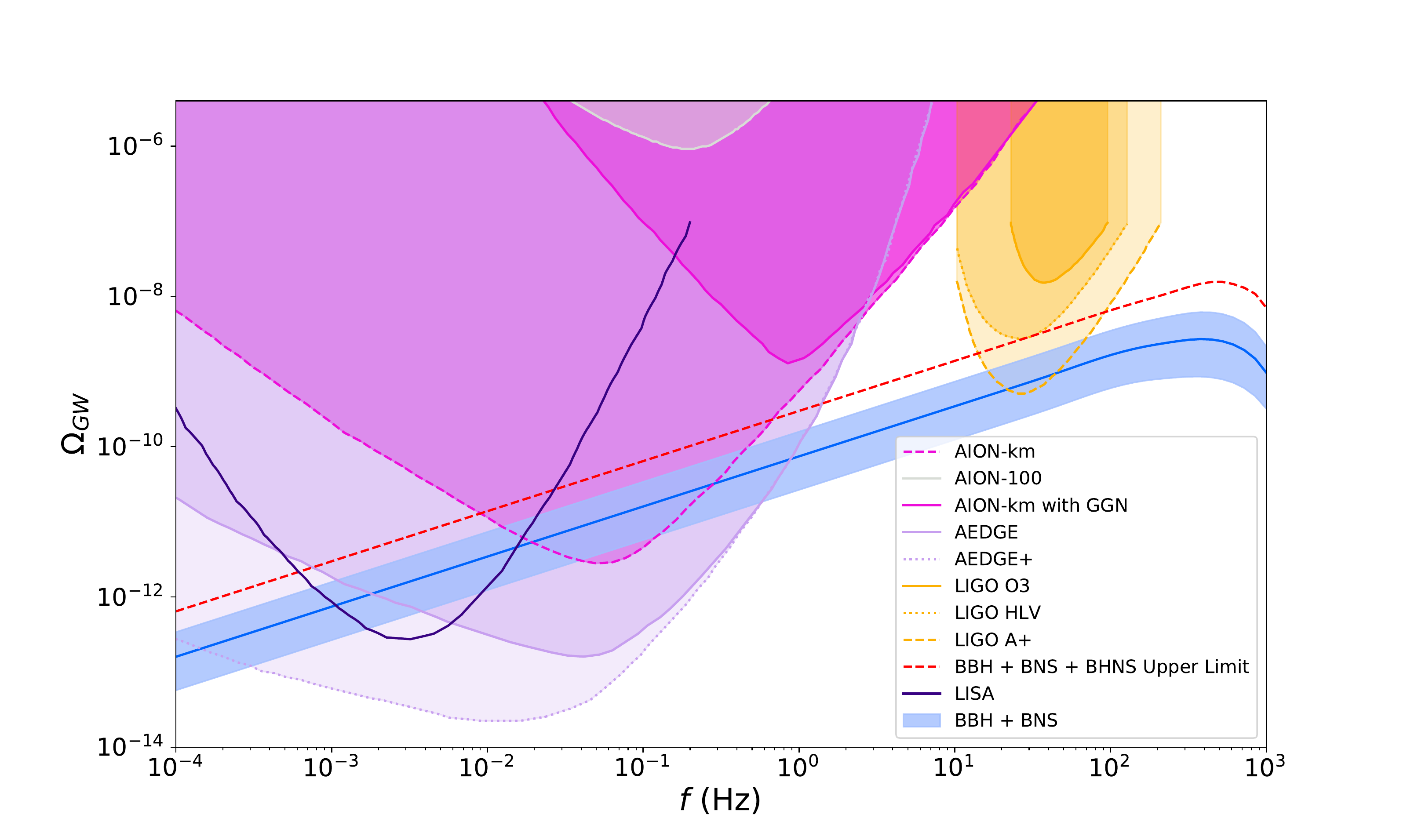}
\caption{The total GWB from BH and NS populations is shown as a solid blue line. The upper (lower) limits of the blue band are linear additions of the upper  (lower) limits of each of these populations individually. The dashed red line is a conservative upper estimate of the total astrophysical signal, obtained by linearly adding the upper limits of each of the BH, NS and BHNS populations. Also shown are the power-integrated (PI) prospective sensitivities to stochastic (used here to mean cumulative or background) signals for the AION-100, AION-km, AEDGE and AEDGE+ experiments taken from Ref.~\cite{Badurina:2021rgt}. We also display the predicted sensitivity for AION-km assuming that gravitational gradient noise (GGN) cannot be eliminated \cite{Badurina:2021rgt}. These estimates are based on Peterson's new low noise model (NLNM) which bounds the expected seismically-induced GGN from below.  The 2$\sigma$ PI curves for the recent third LIGO observation run (O3), as well as projections for the HLV network and A+ detectors operating at design sensitivities are taken from \cite{LIGOScientific:2021djp}, and the prospective sensitivity of LISA \cite{https://doi.org/10.48550/arxiv.1702.00786} to stochastic signals is taken from Ref.~\cite{Renzini:2022alw}.}
\label{fig:aion}
\end{figure}

\subsection{Analysis}
\label{analysis}
In Fig.~\ref{fig:components} we plot our estimates of the GWBs from each of the estimated populations. We show $\Omega_{BBH}$ and  $\Omega_{BNS}$ as bands whose outer limits denote our predictions using the upper and lower bounds of the present-day merger rate. Being a conservative upper limit, $\Omega_{BHNS}$ is shown as a dashed line. We note that the contributions share a common gradient over much of the spectrum. This reflects the fact that much of the gravitational radiation is emitted during the inspiral phase, where $\Omega \sim f^{2/3}$ independent of the nature of the binary system. In Fig.~\ref{fig:aion}, we combine these distinct contributions in order to evaluate the overall implications of the total astrophysical background for terrestrial long baseline atom interferometers. The blue line shows the addition of the central estimations of the emission from BBH and BNS populations, with the surrounding band encompassing the signals obtained from linearly adding the estimates using either the upper or lower values of the present merger-rates of these two  populations. The red dashed line shows the addition of the upper limits from all three populations, giving a somewhat conservative estimate of the maximum astrophysical signal expected in this frequency band. Also shown on this plot are the power-integrated (PI) prospective sensitivities to stochastic signals  for the AION-100,\footnote{This is planned as an intermediate stage between the AION-10 and AION-km experiments. Although more suited to searches for Ultra-Light Dark Matter than GW, we include it here for reference. } AION-km, AEDGE and AEDGE+ experiments taken from Ref.~\cite{Badurina:2021rgt}, where the use of `stochastic' here should be treated as synonymous with cumulative or `background'. We also display the predicted sensitivity for AION-km assuming that gravitational gradient noise (GGN) cannot be eliminated \cite{Badurina:2021rgt}. This estimation deploys Peterson's new low noise model (NLNM) which places bounds on the expected seismically-induced GGN from below. Recent work in Ref. \cite{Badurina:2022ngn} is optimistic of largely mitigating such effects in searches for Ultra-Light Dark Matter, by operating in a multi-gradiometer configuration, although further studies are needed to fully assess the potential for GW detection. For comparison, we also display the 2$\sigma$ PI curves for the recent third LIGO observation run (O3), in addition to projections for the HLV network and A+ detectors operating at design sensitivities \cite{LIGOScientific:2021djp}. The prospective sensitivity of LISA  \cite{https://doi.org/10.48550/arxiv.1702.00786} to stochastic signals, taken from Ref.~\cite{Renzini:2022alw}, is also shown. It is clear that this astrophysical signal would be well within the reach of the terrestrial AION-km experiment (assuming complete mitigation of GGN), its prospective space-based successors, in addition to both LISA and future generations of LIGO. 

Although falling beyond the reach of current GW observatories, (e.g. LIGO O3), the astrophysical background is a well-motivated future target providing complementary information to that obtained from individual merger events. By giving access to objects at higher redshifts than those in resolved detections, studies of GWBs enable the investigation of the collective properties of BH and NS populations such as their average masses, binary occurrence rate \cite{Regimbau_2011} and how these evolve with redshift. GWB studies have also been proposed as a means to extract information on  BH angular momentum, NS ellipticity \cite{De_Lillo_2022} and  NS magnetic fields.  The opportunity to extract information pertaining to stellar formation rates and the evolution of stellar metalicities with redshift is another strong motivation of relevance to a number areas of astrophysics and cosmology. Theoretical studies have also looked to stochastic signals in a bid to investigate hypothetical scenarios such as multi-channel astrophysical and primordial BH mergers \cite{Bavera_2022}. If the subdominant cosmological (primordial) stochastic background is to be observed in the future, as has been identified as a major scientific goal of LISA \cite{https://doi.org/10.48550/arxiv.1702.00786},  this will likely demand some subtraction of the stochastic astrophysical signal. By giving access to this signal in a complementary frequency range to other experiments, atom interferometers should allow for a more complete characterisation and understanding of the spectral shape of the astrophysical GWB. In this way they may play a crucial role in the future unveiling of the cosmological stochastical background.

\section{The ECO Zoo}
\label{ECOS}
Let us now consider the dark sector possibilities which could give rise to a SGWB from mergers of dark compact objects.  We commence with some words of motivation in Sec.~\ref{sec:mot}, however one can skip directly to Sec.~\ref{sec:estimates} for primarily quantitative aspects.

\subsection{Motivation}
\label{sec:mot}
The Standard Model (SM) of particle physics is by no means minimal.  At the hitherto smallest explored distance scales nature exhibits four types of gauge force, a sm\"org\aa sbord of matter fields in a variety of gauge representations with masses spanning at least twelve orders of magnitude and, inexplicably, a solitary scalar field.  Given this diverse offering of microphysics it is not surprising that the SM gives rise to a cornucopia of naturally occurring phenomena over an extraordinary range of scales.  At the upper end one has composite quasi-stable astrophysical objects; a family of star varieties, white dwarfs and neutron stars, all of whose stability and formation are owed to the diverse length scales embedded within those unexplained fundamental SM parameters.  A Universe without the massless photon, without the electron mass endowed by its interaction with the Higgs field, without the proton mass endowed by dimensional transmutation in QCD, or without the tiny proton-neutron mass splitting endowed by the Higgs field and the quark charges, and so forth, would be very different indeed.  Perhaps unimaginably so.

Given the scant information we have on the microphysics of the dark sector and the richness we observe in the visible sector it would thus seem na\"ive to suppose that the dark sector ought to be any less rich.  We do know that the majority of the dark matter budget is not subject to strong long-range dissipative forces, but that is a very weak restriction when it comes to the potential variety of dark phenomena.

What of Occam's Razor?  There are many instances in the past in which Occam's Razor, as interpreted as implying some form of `minimality' of microphenomena which counts particles, would have retrospectively failed.  For instance, the atomic nucleus has, at various stages of our understanding, been a heavy charged sphere, a composite of protons and neutrons, and a composite of composites of quarks and gluons.  As far as counting particles goes, it would be a flagrant violation of Occam's Razor to invoke the quarks and gluons to explain Rutherford's scattering measurements when a simple, heavy charged point particle suffices.  Yet, the quarks and gluons and their associated longer range phenomena are the truth of the matter.  There is, however, reason in a view of Occam's Razor which suggests that one should work with the appropriate effective field theory at the length scales of relevance.  An effective field theory contains only the degrees of freedom relevant at a given length scale.  For example, an effective field theory containing only electrons and a heavy charged nucleus goes a long way if one is only interested in physics at atomic scales and longer, so why worry about the nitty-gritty of QCD?

The problem with the dark sector is that we do not know at which length scale the evidence of microphysics will show up.  Particularly, at which `step' in the potentially rich hierarchical layers of effective descriptions of dark phenomena.  If there is a similar degree of richness in the dark sector as in the visible sector then quasi-stable states may exist over a great range of scales and there are no guarantees as to which states will reveal themselves first.  Furthermore, non-gravitational interactions between the dark and visible sectors have been pursued relentlessly over great ranges of scales in an experimental programme which is consistently revitalised and extended through novel detection strategies and technological developments.  However, what if a rich dark sector exists, but effectively with only gravitational visible sector interactions?  After all, thus far all evidence for the dark sector has been through its gravitational influence on visible matter, hence it would not be entirely surprising if this remained the case.  How, then, might we reveal the richness of the dark sector?

As discussed in the Introduction, the dawn of gravitational wave astronomy has opened new eyes onto the dark sector.   We may even now hope to explore the richness of the dark sector through gravity alone.  Were gravitational wave phenomena directly associated with the physics of fundamental fields detected then this would correspond to Compton wavelengths ranging from $\sim 10^{-25} - 10^{-12}$ eV.  This renders macroscopic condensates of light axion-like fields in this mass range as an interesting particle physics candidate for exotic dark sector phenomena, with rich prospects (see e.g. \cite{Arvanitaki:2009fg,Arvanitaki:2016qwi}), especially if such a wavelength is resonant with existing astrophysical objects such as black holes.  Alternatively, given a sufficiently large degree of red-shifting, gravitational waves in this frequency range could be sourced by physics at much smaller distances which have, subsequently to the dynamical process, undergone cosmological red-shifting.  For example, early Universe phase transitions or microscopic inflationary dynamics could both have undergone significant red-shifting (see e.g. \cite{Caprini:2018mtu} for a comprehensive review).

Finally, in the SM, thanks to the universal attraction of gravity, we also have the aforementioned macroscopic objects comprised of a large number of fundamental constituents held together by gravity.  If such objects exist in the visible sector, then why not in the dark sector?  If such exotic compact objects (ECOs \cite{Macedo:2013jja,Giudice:2016zpa,Cardoso:2016oxy}) exist then they could lead to a wide array of gravitational wave phenomena.  One possibility is that one may have binary ECOs which merge giving rise to a source of gravitational waves (see e.g. \cite{Cardoso:2019rvt} for a comprehensive review).  In this case, what should the characteristic frequency of such mergers be?

Given a dark sector which has its own interactions and is comprised of particles at some mass scale, we should ask if there is an upper limit on the frequency of gravitational waves that could be sourced by ECOs formed of these particles?  When binary orbiting ECOs are far apart the frequency of gravitational waves may be arbitrarily low.  However, as they approach one another and ultimately merge the maximal frequency emitted will parametrically correspond to the merger frequency, which is effectively set by the radius of the objects before they touch.  Thus the ECO radius is the parameter of reference.  If the ECOs are more compact than their event horizon they are necessarily a black hole, thus the minimal size limit for an ECO, and hence maximal frequency of GWs, is essentially the Schwarzschild radius
\beq
R_{\text{Min}} = \alpha 2 G M~~,
\eeq
where $\alpha$ is some constant satisfying $\alpha>1$ which is likely, at least, to be an $\mathcal{O}(1)$ quantity.  The visible sector already provides a guide where $\alpha\gg1$ for a white dwarf and, in practise, $\alpha\sim \mathcal{O}(2-4)$ for a neutron star.  Thus to determine the characteristic GW frequency it remains to determine the characteristic value of $M$, or at least the range of masses possible.

\subsection{Maximal ECO Masses}
\label{sec:estimates}
Consider an effective theory containing particles of mass $m$.  These particles have gravitational interactions and, potentially, additional gauge or self-interactions characterised by a gauge coupling $g$ or scalar self-coupling $\lambda$.  Suppose we start with one particle at rest and add another particle.  If gravity is the only interaction, or if it can overcome the mutual repulsion of other forces, then a composite object forms.  Now consider continuing, adding particles ad nauseam.  If there exists a maximum mass configuration beyond which the composite object collapses to a black hole, or some other more compact object, then what is the mass of this extremal compact object?

We may estimate this mass using dimensional analysis.  We work in the same units as, for instance \cite{Schwinger:1948yk}, where the dimensionless fine-structure constant is $\alpha = e^2/4 \pi \hbar c$, with the sole exception being that we will additionally set $c=1$ as it offers no advantage to carry it through the equations.  In these units, comparing with Newton's law of gravitation we see that the dimensions
\beq
\left[ e^2 \right] = \left[ G_N m^2 \right]  ~~,
\eeq
match, where $G_N$ is Newton's constant.  Thus we see that the parameter combination $G_N m^2 / \hbar$ is dimensionless.

Any physical mass limit can only be expressed as a function of the fundamental parameters of the theory.  These parameters include any potential couplings and the mass.  Thus, if such an upper limit exists, then on dimensional grounds it will approximately take the form 
\beq
M_{\text{Max}}^2 \approx c_{ijk} m^2  \left(\frac{\hbar}{G_N m^2} \right)^i \left( \frac{\lambda}{4 \pi \hbar} \right)^j \left( \frac{g^2}{4 \pi \hbar} \right)^k ~~,
\eeq
where we expect only one $(i,j,k)$ would dominate any ultimate expression for the critical mass.  The constant of proportionality is expected to be some numerical factor which is not necessarily hierarchically large or small.  We expect $i\geq0$, since as gravity becomes weaker the object can get more massive before gravitational collapse occurs, and $i\neq0$ since gravity is the only force holding the object together.    Similarly, for repulsive forces we expect $j,k \geq 0$, as more repulsion allows for a more massive object to form without collapsing.

One obvious light example is a Planck-scale object existing independent of any supporting force $(i,j,k)=(1,0,0)$
\beq
M_{\text{Max}}^2 \propto \frac{\hbar}{G_N} ~~.
\eeq
Such an object is, however, beyond the validity of the effective theory and no definite statements can be made about it.  The second possibility is $(2,0,0)$,
\beq
M_{\text{Max}}^2 \propto m^2 \left(\frac{\hbar}{G_N m^2} \right)^2 ~~.
\eeq
This is the smallest maximal mass one can envisage for a compact object which is supported against gravitational collapse and the factors of $\hbar$ indicate that quantum mechanics must play a role.  Indeed, this case simply corresponds to boson stars \cite{Kaup:1968zz}, whose stability against collapse is attributed to the inability to localise the field due to the uncertainty principle.

Now consider configurations which may be supported by a repulsive force which is strong enough to overcome the gravitational attraction and hence block the $(2,0,0)$ case from being stable against explosion due to repulsion.  Assuming a perturbative coupling, which is where we are able to trust the effective theory, the $(2,1,0)$ case is lighter than the $(2,0,0)$ one. Therefore, the next greatest mass is for $(3,1,0)$,
\beq
M_{\text{Max}}^2 \propto \lambda \frac{\hbar^2}{G_N^3 m^4} ~~.
\eeq
Thus one could potentially have a stable object heavier than a standard boson star, supported by a repulsive force.  Interestingly, this is only true if $\lambda \gtrsim m^2/M_P^2$, reminiscent of the weak gravity conjectures.  In any case, this scenario turns out to correspond to the known self-interacting boson star scenario \cite{Colpi:1986ye}.  The gauge case scales in the same way.

Another way in which the $(2,0,0)$ case could be forbidden is due to Pauli blocking.  This does not involve any non-gravitational interactions, thus the obvious next heaviest candidate is the $(3,0,0)$ case,
\beq
M_{\text{Max}}^2 \propto m^2 \left(\frac{\hbar}{G_N m^2} \right)^3 ~~.
\eeq
Indeed, this is none other than the Chandrasekhar limit \cite{Chandrasekhar:1931ftj} and one sees that quantum mechanics plays a significant role here.

This completes our dimensional-analysis-led study of compact object candidate maximal masses, as we have seemingly exhausted the list of potential stabilisation mechanisms.

\subsection{A GWB from ECOs}
\label{sec:sgwbeco}
\begin{figure}
\centering
\includegraphics[width=\textwidth]{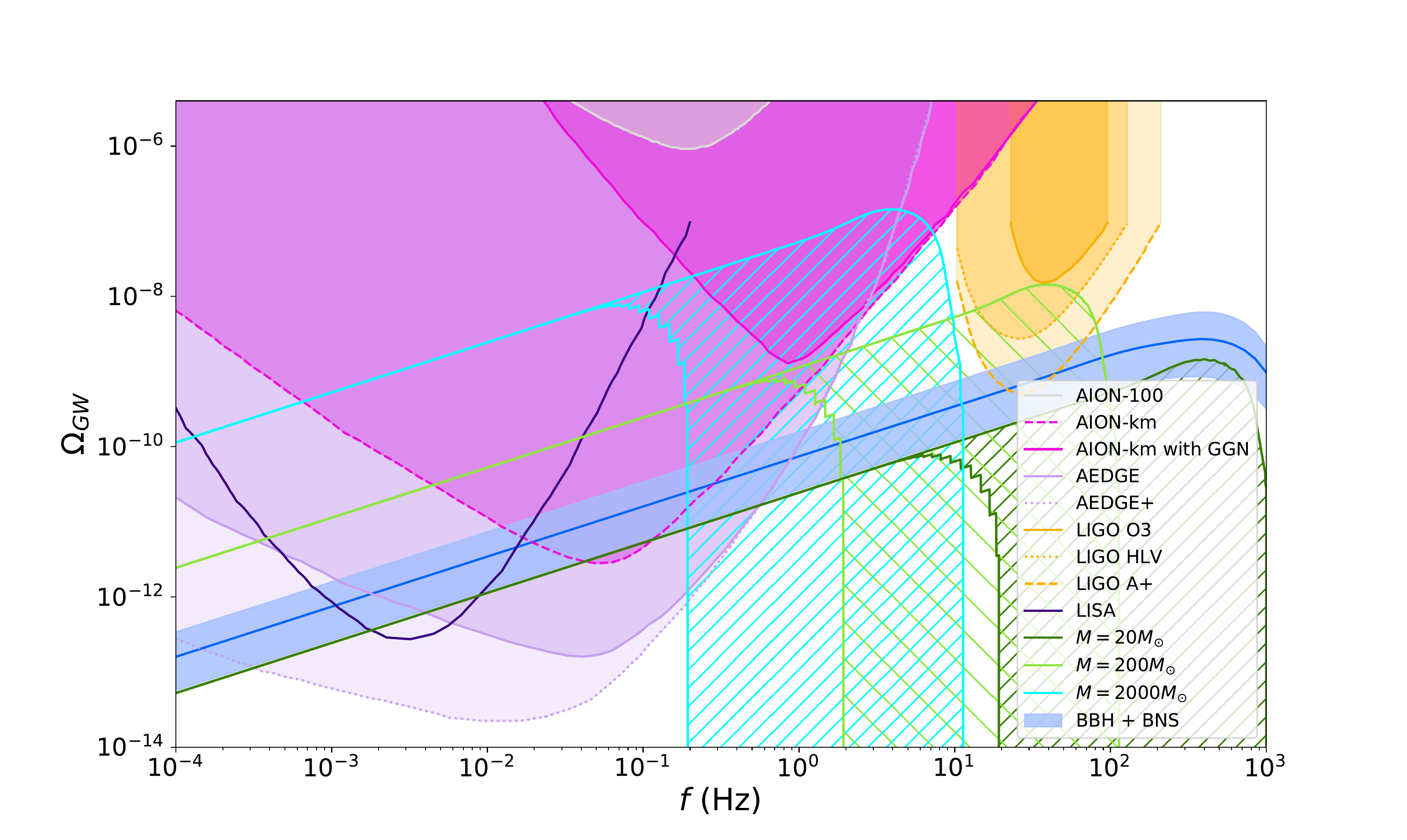}
\caption{An estimation of the potential signal from mergers of populations of  ECOs each of mass $M/2$ where $M = 20, 200, 2000 M_{\odot} $ is the total binary mass, assuming an identical redshift distribution to LIGO black holes and a present merger rate of 10 Gpc$^{-3}$ yr$^{-1}$. The lower frequency cut-off of each band only includes only the inspiral phase, terminating at the ISCO frequency assuming a compactness scale of $C = 0.1$. The higher frequency cut off additionally includes merger and ringdown phases, using the BH wavefunctions given in eq. \ref{eq:wf}.  The cumulative GWB from BH and NS populations is shown as a solid blue line. The upper (lower) limits of the blue band are linear additions of the upper  (lower) limits of each of these populations individually. The dashed red line is a conservative upper estimate of the total astrophysical signal, obtained by linearly adding the upper limits of each of the BH, NS and BHNS populations. Also shown are the power-integrated (PI) prospective sensitivities to stochastic (background) signals for the AION-100, AION-km, AEDGE and AEDGE+ experiments taken from Ref.~\cite{Badurina:2021rgt}. We also display the predicted sensitivity for AION-km assuming that gravitational gradient noise (GGN) cannot be eliminated \cite{Badurina:2021rgt}. These estimates are based on Peterson's new low noise model (NLNM) which bounds the expected seismically-induced GGN from below.  The 2$\sigma$ PI curves for the recent third LIGO observation run (O3), as well as projections for the HLV network and A+ detectors operating at design sensitivities are taken from \cite{LIGOScientific:2021djp}, and the prospective sensitivity of LISA \cite{https://doi.org/10.48550/arxiv.1702.00786} to stochastic signals is taken from Ref.~\cite{Renzini:2022alw}.}
\label{fig:ecos}
\end{figure}

With the various astrophysical populations that could generate a signal within the AION frequency range mapped, and the motivation for the existence of exotic compact objects explored, we now turn to estimating the GWB generated by hypothetical populations of these objects bound in binary systems. Due to our ignorance of the exact nature, specific formation details and distributions of such objects, we will keep our analysis general, parametrising it only in terms of the total mass, $M$, of the binary system.  Given that the GWB is dominated by the inspiral phase in which the waveform is independent of the nature of the objects involved, we will not specify ECO candidates but instead investigate the potential magnitude of signal in terms of the binary mass-scale. In the absence of any empirical knowledge or theoretical motivation for the mass and redshift distributions of these objects, for simplicity we will consider all of the ECOs in a given population to have the same mass, $M/2$.  As we did for BHNS systems, we adopt the same redshift distribution as for stellar-mass BBHs. Whilst the waveform during the inspiral phase is known, the frequency at which this ends depends on the nature of the merging objects, as do the waveforms of the post-inspiral  merger and ringdown phases.  It is however reasonable to consider the end of the inspiral phase to occur when the ISCO is reached. In general, this depends on the compactness, $C$, of the objects involved. Defining this to be the typical mass-radius ratio of the constituent objects in the binary (i.e. $G M/2R$ where $R$ is the typical radius scale), the ISCO frequency as measured at the source is \cite{Giudice:2016zpa}
\begin{equation}
\label{isco}
f^{\textnormal{ECO}}_{\textnormal{I}} = \frac{C^{3/2}}{3^{3/2} \pi G M }~. 
\end{equation}
We note that this encompasses the BH ISCO frequency given in equation \ref{fisco} where the BH compactness ratio is 1/2. This can be taken as an upper limit. In Fig.~\ref{fig:ecos} we plot the resulting signal for various ECO masses taking the present day-merger rate to be $R_0 =$ 10 Gpc$^{-3}$yr$^{-1}$, that of  BBHs. Given the lack of any knowledge of merger and ringdown wavefunctions we plot our estimates as bands: The lower frequency cut-off for each population is where only the inspiral signal is considered, with termination at the ISCO frequency for a value of $C$ = 0.1, approximately that which one might expect for boson stars.  The upper frequency cut-off on our bands includes contributions from inspiral, merger and ringdown phases, employing the BH amplitudes as given in Eq.~\ref{eq:wf}. This likely overestimates the high-frequency spectrum for objects of lower compactness, and can therefore be considered a hard upper limit on the possible high frequency signal.\footnote{Note that the same uncertainties also afflict the interplay between resolved and unresolved background ECO contributions.  Individual ECO mergers would only be resolved if their waveforms, which are likely exotic in the merger and ringdown phases, are actively searched for.  Otherwise, signals with sufficiently large strain may be missed in dedicated searches for BH mergers. In the absence of any guide as to the likely form of these signals, searching for cumulative backgrounds thus provides a robust and general way to probe the presence of such objects across the full range of hypothetical possibilities.}

We see that as the mass scale of the objects increases, the frequency at which the spectrum begins to fall off decreases. In this way, it is possible that there may exist populations of ECOs which produce large signals at AION-km and other atom interferometry experiments, but would not be seen at gravitational observatories with higher operating frequencies such as LIGO. AION thus provides us with an interesting and unique opportunity. On one hand, the observation of  a GWB scaling as $f^{2/3}$, but with a magnitude that differs from what is expected from astrophysical objects only, would indicate the cosmological existence of ECOs. On the other hand, the possible non-observation of this signal at the higher frequencies that are probed in complimentary GW observatories could shed light on the typical  ECO mass scale and, in tandem, the properties and nature of the matter of which they are composed. However, given that backgrounds of IMRIs and EMRIs are highly uncertain, one must not be hasty in concluding that the observation of such a signal is indicative of a population of ECOs. Whilst we must remain open to these more `vanilla' possibilities, current estimates suggest that their backgrounds are, at most, comparable in order of magnitude to that from stellar-mass compact binaries. Any significant mismatch between the well-characterised stellar-mass BBH background and that measured will thus warrant especially close attention, particularly if there are inconsistencies between measurements at different detectors. Until individual waveforms can actually be resolved, or modelling of these astrophysical populations improves, distinguishing between these scenarios could be challenging. Whilst the estimations considered above were made for BH formed from the collapse of baryonic compact objects, they could in principle also form from ECOs. In this sense, the concept of observing an additional background associated with currently poorly understood classes of BH binaries is also somewhat degenerate to observing that from a population of ECOs.

\begin{figure}
\centering
\includegraphics[width=\textwidth]{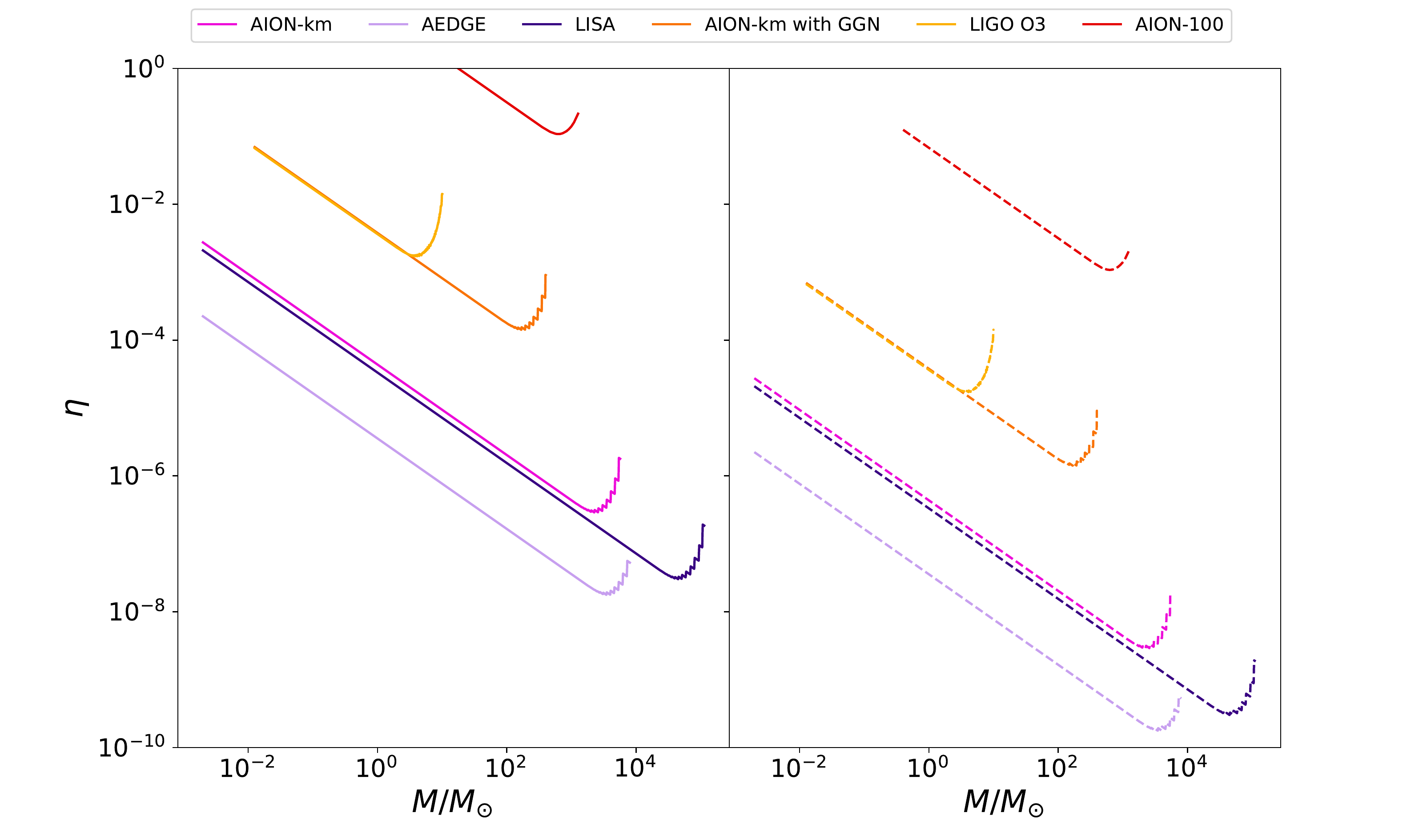}
\caption{A plot of the minimum fraction $\eta$ of dark matter required to be in populations of identical mass binary ECOs as a function of total binary mass $M$, in order for the GWB  to intersect the sensitivity curve of various experiments as shown in Fig. \ref{fig:aion} . The left and right hand panels show the cases where the number of mergers occurring in one Hubble time correspond to 1\%, and 100\% of the total number of binary systems respectively i.e. $\epsilon = 0.01$ and  $\epsilon = 1$.   }
\label{fig:eta}
\end{figure}

In assessing the plausibility of observing backgrounds from ECOs it is also important to be mindful of the mass abundance of ECO binaries that would be required. Assuming that the present merger rate is $R$ Gpc$^{-3}$ yr$^{-1}$ and that a fraction $\epsilon$ of the total number of objects merge within a Hubble time, a population of binary systems of total mass $M$, will harbour a fraction $\eta$ of the total dark matter energy density, $\rho_{\textnormal{DM}} = 0.22 \rho_{c}$
\begin{equation}
\eta = \frac{\rho_{\textnormal{ECO}}}{\rho_{\textnormal{DM}}} \approx  6.4 \times 10^{-7} \times \left( \frac{R}{10}\right) \times \left( \frac{M}{2 M_{\odot}} \right) \times \left(\frac{0.01}{\epsilon}\right)
\end{equation}

\begin{figure}
\centering
\includegraphics[width=\textwidth]{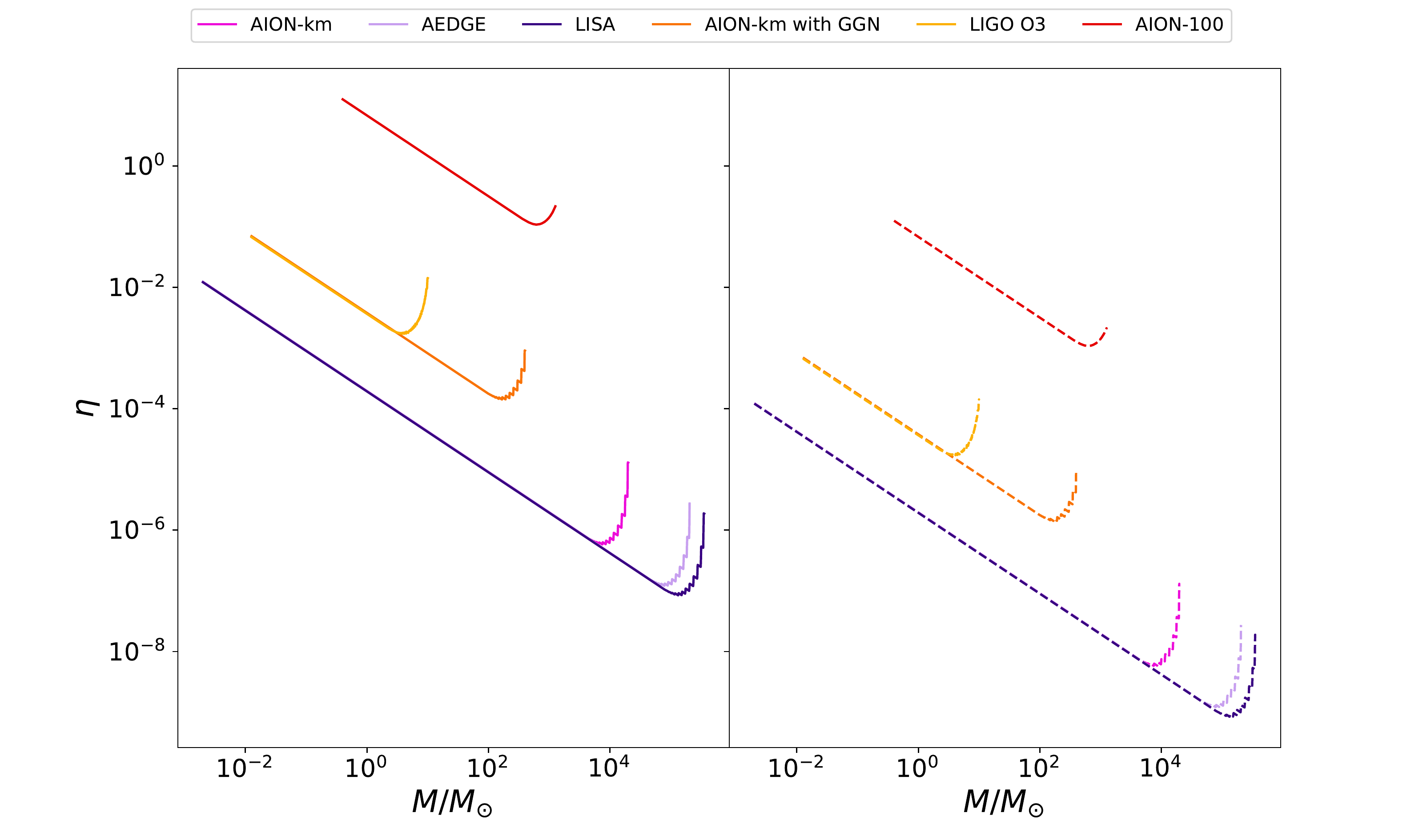}
\caption{A plot of the minimum fraction $\eta$ of dark matter required to be in populations of identical mass binary ECOs as a function of total binary mass $M$, in order for the GWB signal to both intersect the sensitivity curve of various experiments as shown in Fig. \ref{fig:aion} and exceed the cumulative GWB from merging BH and NS (the solid blue line in Fig.~\ref{fig:aion}. The left and right hand panels show the cases where the number of mergers occurring in one Hubble time correspond to 1\%, and 100\% of the total number of binary systems respectively i.e. $\epsilon = 0.01$ and  $\epsilon = 1$.   }
\label{fig:eta2}
\end{figure}

Using $R = 10$ and taking  $\epsilon$ to be 0.01, we find that the populations plotted, which have total masses of $M$ = 20, 200, 2000 $M_{\odot}$,  correspond to $\eta \sim 10^{-5}, 10^{-4}$ and $10^{-3}$ respectively. Hence, even if just a fraction of the dark matter budget is in binaries of ECOs, they could generate a measurable signal in terrestrial long-baseline atom interferometers which could potentially be distinguishable from the expected standard model background. To illustrate this point further, in Fig.~\ref{fig:eta} we plot the value of $\eta$ required for the ECO signal of different mass populations to form a tangent to the minimum of the sensitivity curve of various experiments for both $\epsilon = 0.01$ (left hand panel) and $ \epsilon = 1$ (right hand panel). In Fig.~\ref{fig:eta2} we repeat this analysis now plotting the minimum value of $\eta$ required for the ECO signal to both intersect the sensitivity curve, and exceed the astrophysical GWB which we take to be the sum of the central BH and NS signals (i.e. the solid blue line in Fig. \ref{fig:ecos}).  Both figures corroborate the potential for discovering complexity in dark sector sub-components.

\subsection{Analytic Estimates}
It is useful to gain a qualitative and quantitative understanding of the results presented in Fig.~\ref{fig:ecos} by performing some first-principles estimates, focussing principally on the inspiral-only curves which follow the lower frequency cutoff and are amenable to analytic estimates.

Suppose a fraction $\eta$ of all dark matter is in the form of ECO binaries comprised of equal-mass ECO binaries of total mass $M$.  Furthermore, suppose that a fraction $\epsilon$ of these binaries have merged between formation and now.  We may derive an upper bound on the SGWB density generated by these mergers to be approximately
\beq
\rho_{ECO} \leq \eta \epsilon \kappa \rho_{DM}  ~~,
\eeq
where $\kappa$ is the fraction of rest-mass emitted in GWs for a single binary merger.  Red-shifting will only reduce the right hand side, hence a bound rather than an equality.  It is not possible to accurately estimate $\kappa$ for all stages of a merger, since it depends on the details of the ECO.  However, we may at least estimate the contribution to GWs up to the ISCO point.  The reason is that the change in gravitational potential from the moment the binary forms to the ISCO is
\beq
\Delta V \approx \frac{G M^2}{4 r_{I}} ~~,
\eeq
where $r_{I}$ is the separation of the objects at the ISCO.  Hence the total change in energy from the point in which the ECOs were effectively at infinite separation is
\beq
\Delta E \approx \frac{G M^2}{8 r_{I}} ~~,
\eeq
where the kinetic energy change has also been included.  This energy will have been radiated in GWs, thus allowing an estimate of $\kappa$.  Taking the ISCO separation to be \cite{Giudice:2016zpa}
\beq
r_{I} \approx \frac{3 G M}{2 C} ~~,
\eeq
where $C$ is the compactness ($C=1/2$ for a Schwarzschild black hole) and we have employed the usual Schwarszchild radius, we have
\beq
\kappa \approx \frac{C}{12} ~~.
\eeq
Thus, on relatively general grounds, we expect
\beq
\rho_{ECO} \lesssim \frac{\eta \epsilon C}{12} \rho_{DM} ~~.
\eeq
Now we recall from Sec.~\ref{general} that up until the ISCO the differential GW energy density emitted by a binary scales as $f^{2/3}$.  Thus we may approximate the SGWB for an ensemble of equal mass ECO binaries as
\beq
\Omega \approx \Theta(f_{I}-f) f^{2/3} \Omega_p
\label{eq:ansatz}
\eeq
where $f_{I}$ and $\Omega_p$ are the ISCO frequency and peak dimensionless GW energy density.  Integrating this differential density over frequency up to $f_p$ we thus find that the total GW energy density is related to the peak height as
\beq
\rho_{ECO} = \frac{3}{2} \Omega_p \rho_c ~~.
\eeq
Thus, putting both elements together we may estimate the GW peak height to be
\bea
\Omega & \lesssim &  \frac{\eta \epsilon C}{18} \frac{\rho_{DM}}{\rho_c} \left(\frac{f}{f_I} \right)^{2/3} \\
& \lesssim & \frac{\eta}{10^{-5}}  \frac{\epsilon}{10^{-2}}  \frac{C}{0.1} \left(\frac{f}{f_I} \right)^{2/3} \times 10^{-10}~~,
\label{eq:height}
\eea
where the inspiral-only curves are cutoff at $f_{I}$ which was defined to be
\bea
f_{I} & = & \frac{C^{3/2} c^3}{3^{3/2} \pi G M} ~~,\\
& \approx &  \frac{400 M_\odot}{M} \left(\frac{C}{0.1} \right)^{3/2}~~.
\label{eq:freq}
\eea
Thus, with Eq.~\ref{eq:height} and Eq.~\ref{eq:freq} one can reliably estimate the main features of $\Omega_{GW}$ from ECO binaries.  Indeed, Eq.~\ref{eq:height} appears to be a very good approximation to the peak heights in Fig.~\ref{fig:ecos}.  Thus we see that the estimates provided for the inspiral phase give some physical intuition for the scale of the effect and hence some basic understanding as to why such tiny subcomponents of the dark sector could in principle generate observable signatures in future GW detectors.

\section{Conclusions}
\label{sec:conclusions}
With the advent of gravitational wave astronomy a new kind of light is illuminating the dark.  The opportunities for advancing our knowledge of what has been hitherto hidden from view are substantial.  In this work we have concentrated on observations of a GWB at long baseline atom interferometry experiments.  In Sec.~\ref{general} we have focused on the signatures of known populations of celestial body binaries.  Importantly, we have shown that both future terrestrial and space-based atom interferometers could be sensitive to the GWB produced by binary populations for which the merger rate has already been estimated through observation of binary mergers by the LIGO-Virgo network  \cite{LIGOScientific:2018mvr,  LIGOScientific:2020ibl,  LIGOScientific:2021djp}.  As a result, atom interferometers could play a key role in plugging the frequency gap between Earth and space-based light interferometers, such as LIGO-Virgo \cite{Abbott_2009} and LISA \cite{https://doi.org/10.48550/arxiv.1702.00786}.  Corroborating observations of this relatively well-understood GWB between different experiments operating at different GW frequencies would essentially link our knowledge of binary populations between different stages of binary evolution.  Physically, this would be through a combined measurement of the spectrum amplitude at different frequencies; this would then be compared to the logarithmic slope of the GWB, which is predicated to scale as $f^{2/3}$ below around $100$ Hz.

What if inconsistencies were found between observations at different frequencies?  If such inconsistencies survived scrutiny then they could point to new, unexpected sources of GWBs beyond the Standard Model.  In Sec.~\ref{sec:sgwbeco} we have shown that one potential way in which new signals could show up differently at different detectors is if there is a cosmological population of ECO binaries.  Here, by `exotic' we mean that they are composed of as-yet unknown particles and by `compact' referring to the fact that their radius would be within an order-of-magnitude of the Schwarzschild radius, as for neutron stars.  Interestingly, such ECOs need not comprise the dominant component of the cosmological dark matter density.  Rather, they could form but a tiny fraction, as for neutron stars to the cosmological baryon matter density. Whilst we must remain mindful of a possible degeneracy with backgrounds from currently poorly understood BH populations, the possibility of observing such a signal is certainly an exciting opportunity which could provide an important future probe of the ever-elusive dark sector.

This latter aspect reveals the importance of probing the GWB across a hierarchy of frequencies.  As an example, we have shown that, due to their size, a population of ECOs whose mass satisfies $M \gtrsim 10^3 M_\odot$ would generate a GWB which is all but invisible to the LIGO-Virgo network, but could show up with a pronounced signature at atom interferometers or LISA.  This illustrates the important synergies in probing the GWB across a range of frequencies and highlights the large degree of complementarity between detection technologies.

\acknowledgments
The authors would like to thank Diego Blas, Leonardo Badurina and John Ellis for helpful discussions on the AION project and for comments on this work. We also thank  Marek Lewicki, Ville Vaskonen and Melissa Diamond for consultation on prior work.  This work is supported by the U.S. Department of Energy grant DE-FG02-97ER-41014 (Farrell), and the U.S. Department of Energy, Office of Science, Office of Nuclear Physics and \href{https://iqus.uw.edu}{\color{black}}{InQubator for Quantum Simulation (IQuS)} under Award Number DOE (NP) Award DE-SC0020970. This work is also supported, in part, through the \href{https://phys.washington.edu}{\color{black}}{Department of Physics} and \href{https://www.artsci.washington. edu}{\color{black}}{the College of Arts and Sciences}  at the University of Washington. HB acknowledges partial support from the STFC Consolidated HEP grants ST/P000681/1 and ST/T000694/1 and thanks members of the Cambridge Pheno Working Group for helpful discussions.

\bibliographystyle{JHEP}
\bibliography{biblio}

\end{document}